\documentclass{article}

\usepackage{arxiv}

\usepackage[utf8]{inputenc} % allow utf-8 input
\usepackage[T1]{fontenc}    % use 8-bit T1 fonts
\usepackage{hyperref}       % hyperlinks
\usepackage{url}            % simple URL typesetting
\usepackage{booktabs}       % professional-quality tables
\usepackage{amsfonts}       % blackboard math symbols
\usepackage{nicefrac}       % compact symbols for 1/2, etc.
\usepackage{microtype}      % microtypography
\usepackage{lipsum}
\usepackage{graphicx}
\graphicspath{ {./images/} }
\usepackage{amsmath}

\title{Feedback Dynamics of the Low-Income Rental Housing Market: Exploring Policy Responses to COVID-19}

\author{
 Katherine Marcal \\
  School of Social Work
  Greenspun College of Urban Affairs \\
  University of Nevada Las Vegas\\
  Las Vegas, NV 89154 \\
  \texttt{katherine.marcal@unlv.edu}
\And
 Patrick J. Fowler \\
  Brown School of Social Work, Public Health, and Social Policy\\
  Division of Computational and Data Sciences\\
  Washington University in St. Louis
  \texttt{pjfowler@wustl.edu}
\And
Peter S. Hovmand\\
  Center for Community Health Integration \\
  School of Medicine\\
  Case Western Reserve University
  School of Coumputing and Information\\
  \texttt{psh39@case.edu}\\
}

\begin{document}
\maketitle

\begin{abstract}
The COVID-19 recession threatens mass housing insecurity that undermines economic recovery. Unprecedented federal policy responses halt court-ordered evictions, but questions remain whether policies adequately account for dynamics that drive landlord-tenant interactions including accumulations of rental and mortgage arrears, rental unit availability, and low-income housing options. A system dynamics model probes complex feedback dynamics driving tenant and landlord decision-making in the low-income rental housing market pre- and post-pandemic protections. Feedback loops highlight tradeoffs considered by low-income tenants and landlords in the context of scarcity and uncertainty. Simulations suggest the eviction moratorium reduced evictions by 51\% before its expiration in late 2021, but rental arrears, overcrowding, and homelessness remained high. Federal emergency rental assistance further contributed to keeping evictions low, but showed only modest effects on housing insecurity and homelessness. More rapid allocation of funds would substantially increase impact. Failure to address underlying financial hardship and limited affordable housing undermine COVID recovery.

\end{abstract}

\section{Introduction}
The financial fallout from the COVID-19 pandemic that hit the United States in early 2020 encountered a low-income rental market already marked by instability and scarcity. Housing insecurity and homelessness have long persisted among low-income households in the United States; prior to the pandemic, millions of households were precariously housed, navigating tight low-income rental markets with scarce resources. Extensive shutdowns to mitigate spread of the virus led to an immediate spike in unemployment that impeded low-income families’ abilities to meet basic needs. A federal eviction moratorium passed in spring 2020 aimed to keep people housed during the public health crisis, but did not contain measures to address unpaid arrears that would accumulate in the meantime – thus introducing complex trade-offs and incentives for tenants and landlords. The federal government has since allocated \$50 billion in rental assistance for low-income renters, but complex eligibility requirements and technological limitations have impeded efficient disbursement of funds. In this economic and political context, low-income households have been forced to make difficult tradeoffs about housing costs and quality while landlords navigate unpaid rent, shifting rental market conditions, and frequent policy changes. The complex dynamics driving decision-making among low-income renters and landlords are poorly understood, impeding efficient solutions to persistent housing insecurity. The present study applied system dynamics to examine the feedback mechanisms underlying tenant-landlord decision-making that drive rental market trends. Findings inform understanding of the often-hidden low-income rental housing market, links between eviction and homelessness, and implications for mitigating impacts of the pandemic.

\subsection{Background}
The COVID-19 pandemic triggered widespread financial hardship in a low-income housing market already plagued by insecurity and strain. In 2019, half of households earning up to \$30,000 per year were experiencing severe housing cost burden – defined as paying more than 50\% of household income toward rent – and were thus vulnerable to any economic disruption \cite{joint_center_for_housing_studies_of_harvard_university_americas_2020}. The economic shock of COVID-related shutdowns that began in March 2020 most profoundly impacted low-wage earners in service industries, immediately hindering their ability to pay monthly housing costs. Among households earning up to \$35,000 per year, over half reported a job loss and one in three had missed a rent payment by May \cite{us_census_bureau_household_2021}. By the end of 2020, half of low-income households reported they were likely to be evicted within two months. Loss of income necessary to make monthly rent payments thus devastated the already precarious housing security of low-income families. 

Job losses and COVID-related shutdowns did not correspond to an immediate, dramatic spike in evictions and homelessness, due in part to the federal eviction moratorium that halted processing of most evictions due to nonpayment of rent through summer 2021 \cite{national_low_income_housing_coalition_national_2021}. However, the moratorium allowed unprecedented accumulation of arrears expected to drive a tidal wave of evictions if not paid or forgiven \cite{us_census_bureau_household_2021}. In the post-COVID landscape, households have been forced to weigh ongoing financial strain by accumulating arrears in unaffordable housing, moving in with others to alleviate cost burden while increasing crowding and household conflict, or finding cheaper, poor-quality accommodations. The impact of COVID-19 on the low-income rental market was thus not immediately apparent due to the complex, compensatory processes contributing to highly visible outcomes such as eviction filings and executions, foreclosures, and literal homelessness. 

\subsection{Addressing Complexity in the Low-Income Rental Housing Market}
Complex feedback dynamics drive trends in the affordable housing market that are not fully documented. Few studies have investigated patterns of housing supply and demand, needs, and service use among inadequately housed populations \cite{fowler_solving_2019, marcal_understanding_2021, marzouk_modeling_2016}, and complex decision-making may lead to unintuitive outcomes that are poorly understood. For example, only a portion of evictions lead to literal homelessness, but the impact on frequent moves, doubling up, or other forms of housing insecurity remain largely unseen \cite{desmond_forced_2015,richter_integrated_2021}. Furthermore, informal evictions – resulting from landlord pressure to vacate without going through court proceedings – are not captured by traditional data systems, and thus their impact is difficult to assess \cite{garboden_serial_2019}. Prior computational modeling suggests capacity constraints in the homeless services system shape housing and service-seeking behavior \cite{fowler_solving_2019}, indicating that solely tracking homeless service entries and literal homeless counts will overlook households struggling to navigate the low-income rental market. Additional potential feedback mechanisms that have yet to be investigated include landlord tolerance for late rent given rental market conditions \cite{balzarini_working_2021}, tenant tradeoffs between unaffordability and overcrowding \cite{desmond_forced_2015}, and bottlenecks in eviction processing, service receipt, and filling vacant units \cite{fowler_solving_2019}. The complexity of feedback dynamics in the low-income rental market impede efficient targeting of assistance and development of sustainable strategies to stabilize families.

Tenant and landlord decision-making occurs in the context of limited affordable housing supply and widespread economic hardship, and traditional data systems fail to capture underlying feedback processes \cite{marzouk_modeling_2016}. Low-income households struggle to maintain stable housing in the absence of widely available affordable options. Housing cost burden forces households to live paycheck-to-paycheck, relying on monthly income to afford rent with little to no savings for emergencies \cite{curtis_life_2013,oflaherty_what_2009}. Disruptions to regular paychecks can quickly delay rent payments, risking eviction. Unpaid rent likewise strains landlords in the low-income rental market, who are primarily small “mom and pop” operations with one or two rental properties \cite{coulton_characteristics_2020} Rent payments generate landlord income, which is then used to pay mortgages. Late rent can thus increase foreclosure risk over time, although the foreclosure process moves more slowly and offers more opportunities for borrowers to resolve cases \cite{noauthor_real_2017}. Foreclosures have immediate implications for landlords, who lose income from rental units; foreclosures also impact tenants residing in foreclosed units, who are displaced. Finally, foreclosures impact the market more broadly by removing available affordable rental units, constraining supply.

The complex dynamics of the low-income rental market impede effective, efficient policy solutions to evictions, scarce affordable housing, and homelessness. Tracking rates of evictions, foreclosures, and homelessness alone overlook the complex decision-making processes that drive outcomes. Actors alter their behavior in response to others’ behaviors as well as market conditions. Reinforcing processes represent vicious or virtuous cycles that compound system outcomes over time; in contrast, stressors to the system trigger compensatory processes that undermine runaway growth. The economic shock triggered by the pandemic illuminates these dynamics by disrupting business as usual, thus providing a crucial opportunity to understand the low-income rental market and develop effective, sustainable solutions to housing insecurity and homelessness. 

Prior research applying system dynamics modeling shows promise for illuminating the complex feedback processes driving behavior. Studies investigating the Dutch \cite{eskinasi_houdini_2011} and Egyptian \cite{marzouk_modeling_2016} housing markets highlight the roles of construction, zoning policies, and broader economic context in shaping housing options, which are most limited for low-income populations \cite{schuetz_improve_2020}. Policy barriers that impede increasing the affordable housing supply contribute to ongoing instability for low-income households. System dynamics has further been applied to investigate the needs and service use patterns of homeless households \cite{fowler_solving_2019,marcal_understanding_2021}. Failure to account for difficult tradeoffs made in the face of scarce resources risks exacerbating housing insecurity that goes largely unseen and unaddressed.

The following paper uses system dynamics simulation modeling to address the following research questions: (1) What are the main tenant and landlord feedback dynamics driving trends of evictions, foreclosures, and homelessness in the low-income rental housing market? (2) What are the projected effects of the COVID-triggered economic fallout and federal eviction moratorium on these dynamics? Simulation modeling addresses data limitations and dynamic complexity in relationships between tenants and landlords, evictions and foreclosures, and housing insecurity and homelessness. Findings will improve understanding of the complex dynamics driving tenant and landlord decision-making as well as long-term insights into addressing housing insecurity for low-income households.

\section{Model Structure}
A stock-and-flow structure models the low-income rental housing market. The structure draws upon prior empirical research, theory, and consultation with housing and homelessness experts. The model is formulated and simulated according to a series of integral equations to track the dynamics of rent and mortgage payments, eviction, turnover, and homelessness among low-income renters, landlords, and rental units. The model includes four primary substructures: tenant rent payments, landlord mortgage payments, rental units, and tenant households. Given focus on eviction and homelessness risk, the model population is limited to low-income, housing insecure renters – defined as those who earn less than 130\% of the Federal Poverty Level and cannot maintain housing without risk of experiencing housing cost-burden, doubling up, or eviction. Rental units are limited to those that are low-cost and occupied by or available to low-income renters. When tenants experience income increases that substantially alter their ability to afford housing, they are considered to stabilize and thus transition outside of the model boundary.

\subsection{Rent Payments}
Rent payments become due each month (R{d}) according to the average monthly rent per unit [dollars/rental unit/month] multiplied by total occupied rental units in the system (Figure A1). Total rent due is calculated as an accumulation of rent becoming due each month given payments  (Equation 1), and accumulates until it is paid off (R{P}; Figure 1). The outflow R{P} depends on the average time for residents of a rental unit to pay rent (Equation 1.1).

Equation 1:
\begin{equation}\label{eq1}
\begin{split}
R= \int_0^t(R_d-R_p )dt+R_0 \\
Where R = Total rent due [dollars] \\
R_d = Monthly rent becoming due [dollars/month] \\
R_p = Rent payments [dollars/month] \\
\end{split}
\end{equation}

Equation 1.1:
\begin{align}%\label{eq1.1}
R_p =\frac{R}{AT_r}\\
Where: &AT_r = Average time for tenants to pay rent
\end{align}

\begin{figure} % picture
    \centering
    \includegraphics{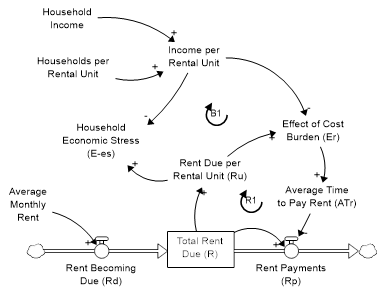}
    \caption{Rent payments accumulating due to housing cost burden}
\end{figure}

AT\_r increases (delaying rent payments) as the ratio of rent due to household income exceeds 0.30 – a standard threshold above which households are considered “cost-burdened” by housing (Equation 2). This delay occurs through an effect (E\_r) fit according to an asymptotic curve (Figure 2). Thus if we assume the initial average time to pay monthly rent is one month, the actual average time is one month multiplied by an effect that begins to increase asymptotically when rent due exceeds 30\% of income per rental unit, with E\_R = 1 indicating no effect. A MAX function prevented the value of the equation from going below 1, indicating lack of problematic household economic stress and thus no effect on time to pay rent. Given the model included only low-income, housing insecure households typically experiencing high housing cost burden, it was assumed that tenants would not pay ahead on their rent \cite{joint_center_for_housing_studies_of_harvard_university_americas_2020,urahn_household_2016}. This mechanism may function differently for higher-income households. Rent due (R) accumulates unless paid by the tenant or forgiven. A reinforcing loop (R1) describes how inability to afford rent each month leads to persistent late payments and accumulating arrears. 

\begin{figure} % picture
    \centering
    \includegraphics{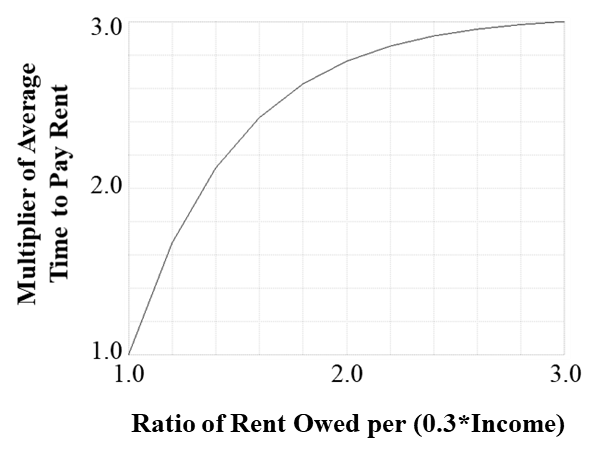}
     \caption{Graphical function linking housing cost burden to time to pay rent} 
\end{figure}

Household economic stress reflects the strain on household finances given housing costs; as the accumulated arrears per household increase relative to new monthly rent due, stress increases asymptotically, driving household decisions to move or double up in order to avoid eviction or homelessness (Figure 3). Household economic stress acts as a multiplier on the baseline turnover rate with E\_es = 1 indicating no effect (Equation 3). We assume exceeding the rent burden threshold quickly impacts ability to pay rent as basic and thus use an asymptotic curve to represent the effect. Values do not fall below (1,1) as it is assumed tenants live on the edge of affordability and do not pay “ahead” on rent.

\begin{figure} % picture
    \centering
    \includegraphics{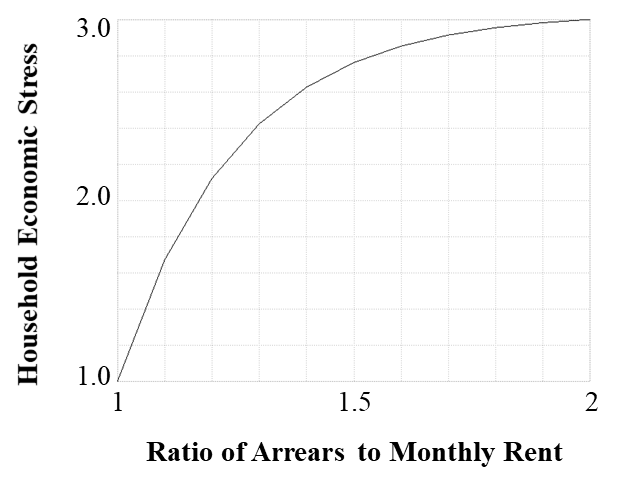}
     \caption{Graphic function linking arrears with household economic stress} 
\end{figure}

Equation 3: 
\begin{equation}%\label{eq3}
\begin{split}
E_es = y_f + (y_0 - y_f) * exp(-exp(alpha)*t) \\
Where:		y_f = asymptote = 3.0 \\
y_0 = y-intercept = -108.2 \\
\alpha = growth rate = 1.4 \\
\end{split}
\end{equation}

Graphical functions were shaped according to the theoretical assumption that late payments have the largest impact early on, as households transition from being tenants in good standing to officially delinquent and thus at risk of losing housing, with this impact tapering off with increasing debt. We assume no additional fees or secondary borrowing associated with late rent. Equations were fit to the points generated by graphs.

\subsection{Mortgage Payments}
Like rent, mortgage payments come due each month (M\_d; Figure 4). It is assumed that landlords rely on income generated from tenant rent payments (R\_p) to make mortgage payments (M\_P); when unpaid rent accumulates, landlord income decreases and mortgage likewise accumulates (Equation 4).

\begin{figure} % picture
    \centering
    \includegraphics{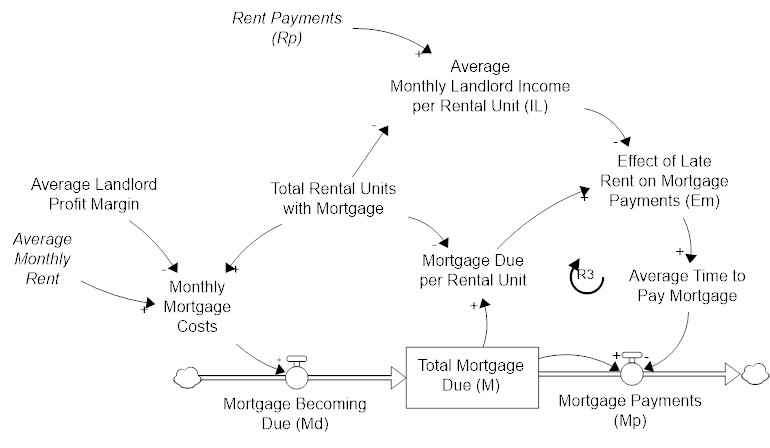}
     \caption{Mortgage payments delayed due to reduced landlord income from rent payments} 
\end{figure}

In a process analogous to tenant income impacting ability to pay rent, decreased landlord income due to missed rent payments (R\_p) delays mortgage payments (M\_P). When mortgage due (M) exceeds income per rental unit generated for landlords (I\_L), mortgage payments are delayed by a multiplicative effect (E\_m; Equation 5) generated by an S-shaped curve with E\_m=1 indicating no effect (Figure 5). Thus, mortgage due (M) accumulates unless paid by landlords. The S-shaped effect reflected the assumption that landlords would have more resources than tenants, and thus experience strain from overdue payments more slowly than would low-income tenants.

\begin{figure} % picture
    \centering
    \includegraphics{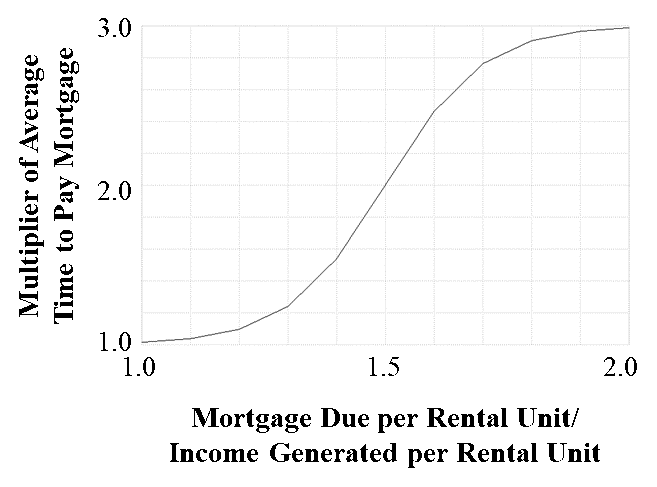}
     \caption{Graphical function linking mortgage due to time to pay mortgage} 
\end{figure}

Equation 5:	
\begin{equation}\label{eq5}
\begin{split}
E_m=Y_(Em,max)+  \frac{(Y_(Em,min)-Y_(Em,max))}{(1+(\frac{r_Em}{e_Em} )^(b_Em ) )}  \\
Where:		Y_(Em,max) = upper asymptote = 3.0\\
Y_(Em,min) = lower asymptote 1.0\\
e_Em = inflection point = 1.5\\
b_Em = slope = 10.0\\
r_Em = M/I_L  = ratio of mortgage due to landlord income per rental unit\\
\end{split}
\end{equation}

\subsection{Rental Units}
A third stock-and-flow structure captures low-cost rental units that may transition through four states: occupied, pending eviction, unoccupied, and foreclosed (Figure 6). Equations 6-6.3 describe values of each state over time. Occupied rental units are occupied by tenants in good standing and up-to-date on rent payments. Rental units pending eviction are occupied by tenants against which an eviction has been filed but not yet processed. Unoccupied rental units are those available for rent but currently unoccupied. Foreclosed units are rental units that have been removed from the rental market with occupants removed as a result of foreclosure.

\begin{figure} % picture
    \centering
    \includegraphics{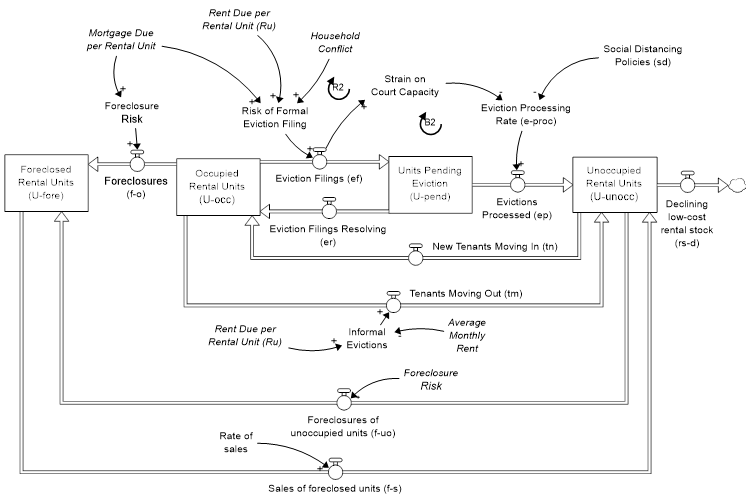}
     \caption{Rental units becoming unoccupied due to evictions and mobility; foreclosures reducing affordable housing stock} 
\end{figure}

Equation 6:	
\begin{align}
U_occ= \int_0^t(t_n+e_r- t_m-f_o- e_f )dt+U_(occ,0)
\end{align}
Equation 6.1:		
\begin{align}
U_pend= \int_0^t(e_f- e_p- fS_r )dt+U_(pend,0)
\end{align}
Equation 6.2:		
\begin{align}
U_unocc= \int_0^t(e_p+ t_m+f_s-t_n-f_uo- s_d )dt+U_(unocc,0)
\end{align}
Equation 6.3:		
\begin{equation}
\begin{split}
U_fore= \int_0^t(f_o+ f_uo-f_s )dt+U_(fore,0)\\
Where:		U_occ  = Occupied rental units [rental units]\\
U_pend  = Rental units pending eviction [rental units]\\
U_unocc  = Unoccupied rental units [rental units]\\
U_fore = Foreclosed rental units [rental units]\\
e_f = Eviction filings [rental units/month]\\
e_r = Eviction filings resolving [rental units/month]\\
e_p = Evictions being processed [rental units/month]\\
t_m = Tenants moving out [rental units/month]\\
fo = Foreclosures of occupied units [rental units/month]\\
fuo = Foreclosures of unoccupied units [rental units/month]\\
f_s = Sales of foreclosed units [rental units/month]\\
t_n = New tenants moving in [rental units/month]\\
rs_d = Declining low-cost rental stock\\
\end{split}
\end{equation}

Accumulated unpaid rent triggers eviction filings (e\_f) through an effect E\_or (Equation 6.4). A minimum constraint on E\_or is set to 1 such that when accumulated unpaid rent exceeds monthly rent owed, risk for eviction filing increases; when accumulated unpaid rent is less than rent owed, E\_or=1 indicating no effect. Eviction risk similarly increases when landlords feel financial pressure, as they rely on rent income to make mortgage payments; landlords are more likely to file evictions when they fall behind on mortgage payments. Finally, eviction risk is also driven by household conflict (discussed in the following section); overcrowded housing can increase charges of excessive noise and nuisance that lead to eviction filings.

Equation 6.4:
\begin{equation}%\lablel{eq6.4}
\begin{split}
E_or=  \frac{R_u}{L_tol} \\
Where:		R_u = Accumulated unpaid rent per unit [dollars/unit]\\
L_tol = Landlord tolerance for overdue rent [dollars/unit]
\end{split}
\end{equation}

Once filed, evictions must be processed. The processing rate is a function of housing court capacity; prior to the pandemic, courts processed approximately 38\% of eviction filings per month. Courts experience strain and slowed processing when a surge in eviction filings occurs. COVID-19 social distancing policies (sd) have further slowed court proceedings. Thus, the proportion of evictions processed is reduced by strain on the courts as well as social distancing policies; the number of evictions processed per month e\_p) is equal to the proportion of evictions processed divided by the average time to process one eviction (Equation 6.5).

Equation 6.5:
\begin{equation}%\lablel{eq6.5}
\begin{split}
e_p=  \frac{e_prop}{AT_proc} *sd \\
Where:		e_prop = Initial proportion evictions processed per month [dimensionless]\\
			AT_proc = Average time to process one eviction [months]
\end{split}
\end{equation}			
 
\subsection{Households}
The model tracks low-income households at risk for eviction and homelessness (Figure 7). “Housing Insecure Households” includes renter households earning up to 150\% of the federal poverty level who are cost-burdened by housing (paying at least 30\% of income toward rent; Equation 7). “Literally Homeless Households” encompasses households living in shelters, on the streets, in abandoned buildings, in vehicles, or anywhere else not meant for human habitation (Equation 7.1) 

\begin{figure} % picture
    \centering
    \includegraphics[scale=.75]{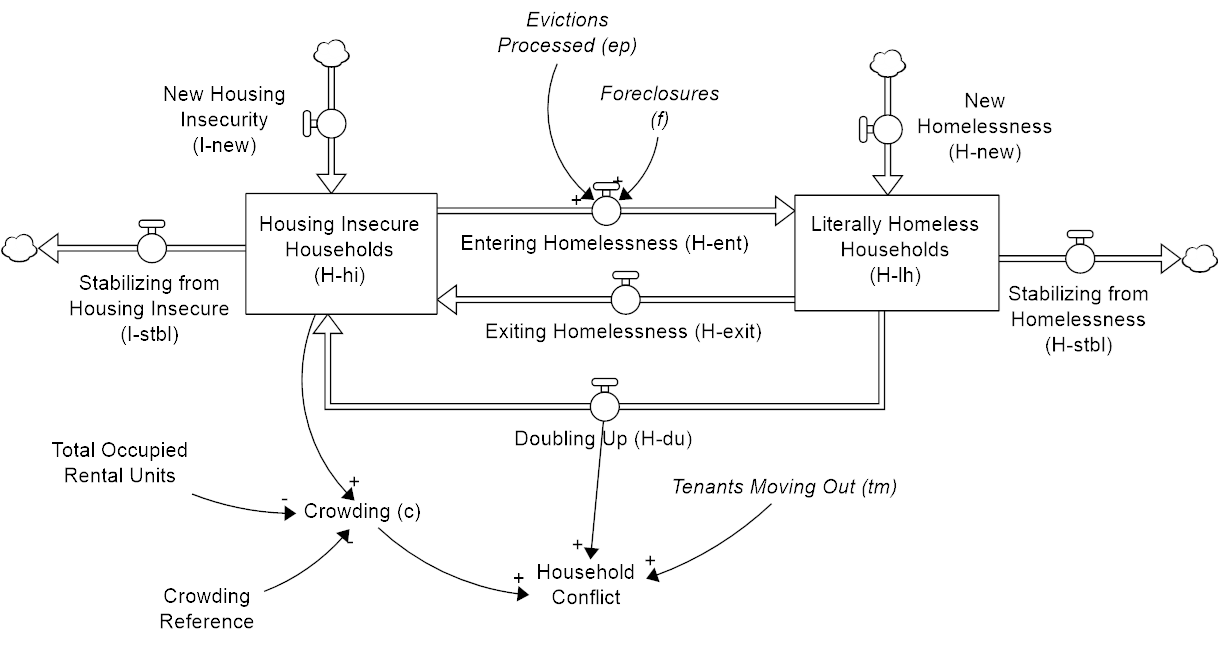}%\includegraphics[width=5cm, height=4cm]{figure_7}%\includegraphics{figure_7.png}
     \caption{Households transitioning between housing insecure and literally homeless} 
\end{figure}

Households may enter either setting for the first time (“New Housing Insecurity” (I\_new)  and “New Homelessness” (H\_new) or stabilize from either setting (I\_stbl and H\_stbl). A proportion of evictions and foreclosures will result in homelessness; however, the majority will lead to families moving into cheaper housing or doubling up, thus remaining in the stock “Housing Insecure Households.” It is assumed that when households are evicted, move to alternate housing, double up, or enter homeless, they do not carry any remaining rent debt with them.

Equation 7:	
\begin{align}
H_hi= \int_0^t(I_new+H_exit+H_du- H_ent-I_stbl )dt+H_(hi,0)
\end{align}

Equation 7.1: 		
\begin{equation}%\lablel{eq7.1}
\begin{split}
H_lh= \int_0^t(H_new+H_ent-H_exit-H_du- H_stbl )dt+H_(lh,0)\\
Where:			H_hi = Housing insecure households [households]\\
I_new = New housing insecurity [households/months]\\
I_stbl = Stabilizing from housing insecurity [households/months]\\
H_lh = Literally homeless households [households]\\
H_ent  = Entering homelessness [households/months]\\
H_exit  = Exiting homelessness [households/months]\\
H_du  = Doubling up after homelessness [households/months]\\
H_new = New homelessness [households/months]\\
H_stbl = Stabilizing from homelessness [households/months]\\
\end{split}
\end{equation}

Crowding (c) results as more households move into a finite number of rental units, thus increasing the number of units housing doubled up households (defined as >1 household per rental unit; Figure 8; Equation 8). Increased crowding drives household conflict via an effect  E\_cr (Equation 8.1). When the average number of households reaches 1 or fewer, there is assumed to be no “crowding” effect.

Equation 8:	
\begin{align}
c=  \frac{[\frac{H_{hi}}{(U_{occ} + U_{pend})}]} { c_{ref} }
\end{align}

Equation 8.1:	
\begin{equation}%\lable{eq8.1}
\begin{split}
E_{cr}=Y_{Ecr,max} + \frac{ Y_{Ecr,min} - Y_{Ecr,max}} {1+(\frac{r_{Ecr}{e_{Ecr}}}) ^ b_{Ecr} }  \\
Where:	Y_(Ecr,max) = upper asymptote = 2.0 \\
Y_{Ecr,min} = lower asymptote = 1.0 \\
e_{Ecr} = inflection point = 1.5 \\
b_{Ecr} = slope = 5 \\
r_{Ecr} = ratio of actual households per rental unit to reference value (1 household per unit)
\end{split}
\end{equation}

Thus, crowding increases as the number of households per total occupied rental units exceeds the crowding reference value (1 household/rental unit). Household conflict was calculated as the product of crowding and turnover effects.
\begin{figure} % picture
    \centering
    \includegraphics{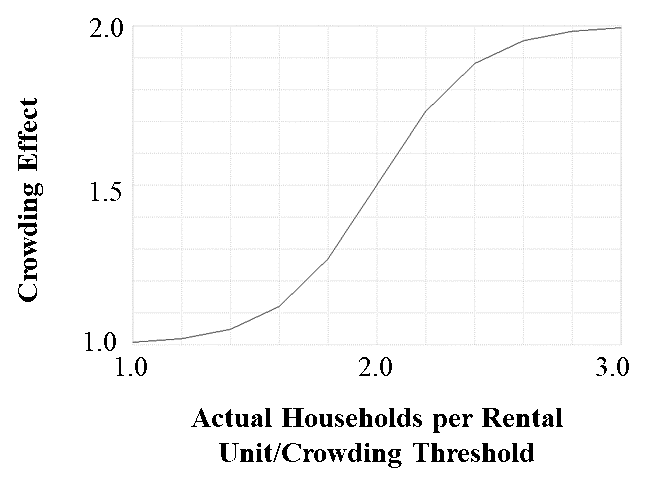}
     \caption{Graphical function linking households per rental unit with crowding effect} 
\end{figure}

\subsection{Key Feedback Mechanisms}
The structures described above are linked by key feedback mechanisms driving tenant and landlord behavior in the low-income rental housing market. Three primary feedback loops describe tenant behavior in response to economic disruption (Figure 9). Timely payments depend upon household income remaining at a level adequate to cover rent payments given other basic needs \cite{balzarini_working_2021}. If household incomes decrease, rent payments are delayen the wake of the COVID-related shutdowns, households face financial strain as they struggle to afford monthly expenses given a period of reduced income. This strain delays rent payments as families must make tradeoffs about which bills to pay first \cite{anderson_coping_2012, beatty_is_2014} thus slowing the outflow of overdue rent and contributing to the accumulation of arrears. Accumulating debt further strains families in a vicious cycle (R1).

\begin{figure} % picture
    \centering
    \includegraphics{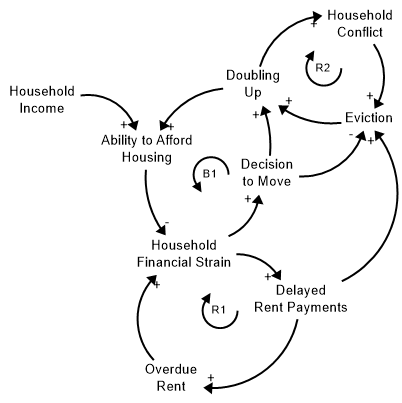}
     \caption{Feedback loops describing accumulation of unpaid rent (R1), tenant tradeoffs in the wake of household financial strain (B1), and consequences of doubling up to alleviate housing cost burden (R2)} 
\end{figure}

A balancing loop describes the tradeoffs tenants face in the wake of household economic distress (B1). Disrupted or inadequate income creates household financial strain as families struggle to afford basic needs. Without action, families continue to accumulate arrears and risk eviction by landlords; conversely, families may choose to leave their current housing knowing arrears can never be paid, and either move somewhere cheaper or double up with friends or family. Leaving unaffordable housing can alleviate financial strain and prevent families from facing formal evictions. 

Third, a reinforcing loop from moving triggers exposure to other stressors such as poor quality, overcrowded conditions and household conflict. Thus, consequences of alleviating financial strain through moving emerge via increased household conflict, which can trigger eviction filings for nuisance-related reasons; abrupt loss of housing can thus drive further need for doubling up as families struggle to find stable, affordable housing in the wake of eviction (R2).

A second set of major feedback loops model landlord decision-making in response to delayed rent payments and market conditions (Figure 10). Most landlords in the low-income rental market are small “mom-and-pop” operations with just one or two units who rely on income from tenant rent payments to afford their mortgages \cite{coulton_characteristics_2020, raymond_foreclosure_2018}. In the wake of major disruption to tenant incomes such as the pandemic-relation shutdowns, a reinforcing loop increases evictions due to landlords’ own financial hardship (R3) while a balancing process slows mass evictions even in the absence of widespread policy protections (Figure 10).

\begin{figure} % picture
    \centering
    \includegraphics{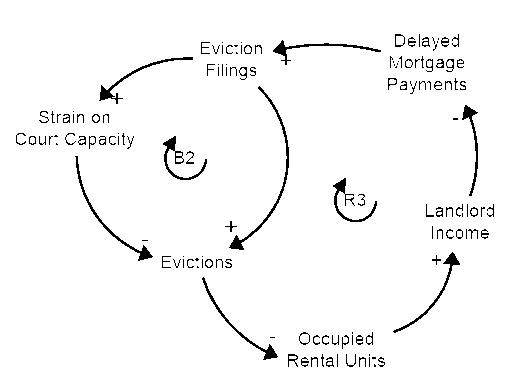}
     \caption{Feedback loops describing fluctuating landlord income as a result in changing rental unit occupancy through evictions (R3) as well as strain on the housing court system slowing evictions as a result of surges (B2)} 
\end{figure}

The reinforcing loop (R3) details the mechanism by which landlords’ own financial pressures drive eviction decisions. When landlords fall behind on mortgage payments, they are more likely to initiate eviction proceedings – either as an attempt to displace a non-paying tenant or pressure tenants into catching up \cite{garboden_serial_2019}. Eviction filings increase processed evictions, which reduce rental unit occupancy and thus landlord income, further delaying mortgage payments. A balancing loop (B2) describes how the volume of eviction filings might impact processing time, thus keeping units occupied. An increase in eviction filings in a period of economic downturn strains court capacity to process cases, thus slowing evictions. This in turn keeps rental units occupied and contributing to landlord income; this delay allows tenants more opportunity to resolve eviction filings by catching up on payments, working out payment plans, or seeking rental assistance. Failure to pay would lead to repeated eviction filings or landlord pressure to vacate voluntarily.

\subsection{Economic Impact of COVID-19}	
The COVID-19 pandemic was modeled using a positive step function representing the overnight exogenous financial shock, while a smoothed negative step function represented a prolonged economic recovery (Equation 14). The magnitude of the spike was calibrated to results of the U.S. Census Bureau’s Pulse Survey launched in April 2020 on household impacts of the pandemic (U.S. Census Bureau, 2021) while the shape of the effect was calibrated to the U.S. unemployment rate (Federal Reserve Bank of St. Louis, 2021).

Equation 11.
\begin{equation}%\lable{eq11}
\begin{split}
STEP(M_e,TS_e )-SMTH(STEP(M_e,TS_e),D_e ) \\
Where 	M_e = Magnitude of economic impact\\
TS_e = Start time of economic impact\\
D_e = Duration of economic recovery
\end{split}
\end{equation}
	
\begin{figure} % picture
    \centering
    \includegraphics[scale=.75]{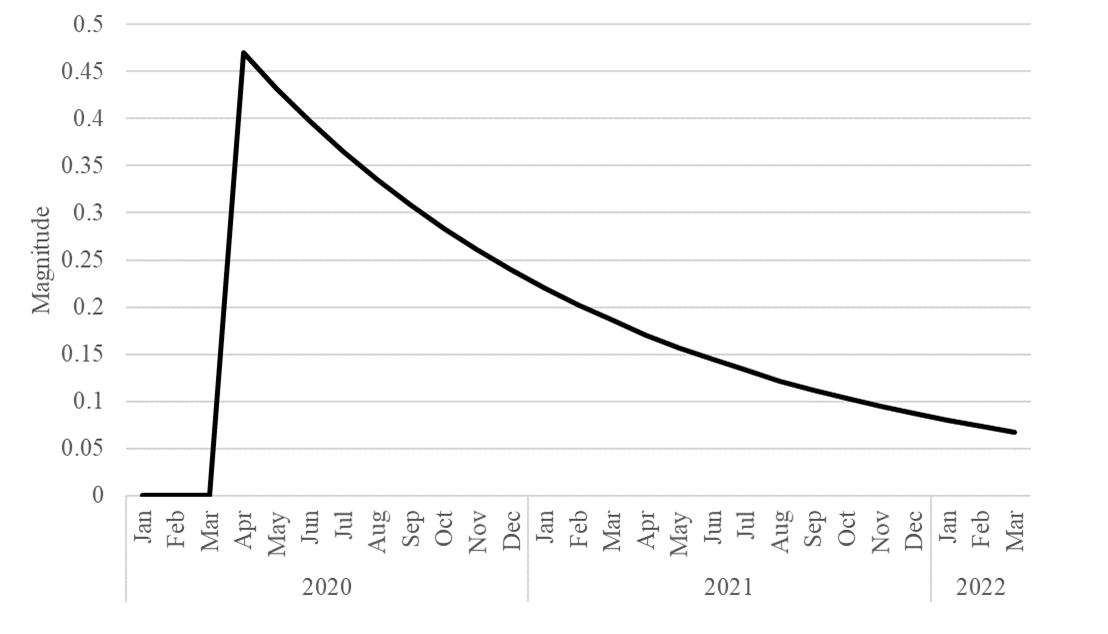}%\includegraphics{figure_11.png}
     \caption{Graphical function representing the economic impact of COVID-19 on low-income households} 
\end{figure}		
		
The economic impact of COVID-19 was included as an exogenous effect at five points throughout the model structure. The effect acted to reduce average household income, increase the rates of new housing insecurity and homelessness, and reduce the rates of stabilization from housing insecurity and homelessness (Equations 11.1-11.5). Additionally, the COVID-19 effect reduced housing court capacity as a result of social distancing guidelines (Equation 14.6), reducing the number of cases that could be processed during the height on the pandemic and acting as a balance on evictions.

Equation 11.1. 	
\begin{equation}%\lablel{eq11.1}
\begin{split}
Average Household Income = \\
INIT Average Household Income * (1 - COVID-19 Economic Impact)
\end{split}
\end{equation}
Equation 11.2.	
\begin{equation}%\lablel{eq11.2}
\begin{split}
Rate of New Housing Insecurity = \\
INIT Rate of New Housing Insecurity * (1 + COVID-19 Economic Impact)
\end{split}
\end{equation}
Equation 11.3. 
\begin{equation}%\lablel{eq11.3}
\begin{split}
Rate of New Homelessness = 
\\INIT Rate of New Homelessness * (1 + COVID-19 Economic Impact)
\end{split}
\end{equation}
Equation 11.4. 	
\begin{equation}%\lablel{eq11.4}
\begin{split}
FR Stabilizing from Housing Insecurity = \\
INIT FR Stabilizing from Housing Insecurity = \\
* (1 - COVID-19 Economic Impact)
\end{split}
\end{equation}
Equation 11.5.	
\begin{equation}%\lablel{eq11.5}
\begin{split}
FR Stabilizing from Homelessness = \\
INIT FR Stabilizing from Homelessness = \\
* (1 - COVID-19 Economic Impact)
\end{split}
\end{equation}
Equation 11.6.
\begin{equation}%\lablel{eq11.6}
\begin{split}
Proportion of Eviction Filings Processed = INIT Proportion of Evictions Processed \\
* Reduced Court Capacity Due to Social Distancing * Eviction Moratorium\\
Where:	Reduced Court Capacity Due to Social Distancing = \\
1 – COVID-19 Economic Impact
\end{split}
\end{equation}

Reduced court capacity to process evictions acts as a balancing force on the number of evictions following a surge in filings, which strains courts and increases processing time.

\subsection{Policy Interventions}
Two additional structures testing impacts of COVID relief policies for renters. First, the federal eviction moratorium enacted in late March 2020 halted most evictions due to nonpayment of rent, which comprise the majority of evictions in the U.S. each year. The moratorium was extended several times before expiring in most places by Fall 2021 \cite{mccarty_federal_2021,national_low_income_housing_coalition_national_2021}. A switch turned the moratorium on in the third week of March 2020, reducing the proportion of evictions processed by 90\% for 18 months (Equation 12). An additional effect likewise reduced eviction filings, reflecting nationwide data suggesting landlords were slow to resume filings during and after the moratorium (Equation 12.1). 

Equation 12. 
\begin{equation}%\lablel{eq12}
\begin{split}
Eviction Moratorium = -STEP(INIT_Proportion_of_Evictions_Processed\\
*ES_EM*Eviction_Moratorium_ON, TS_EM) \\
+ STEP(INIT_Proportion_of_Evictions_Processed*ES_EM*\\
Eviction_Moratorium_ON, ET_EM)
\end{split}
\end{equation}

Equation 12.1.
\begin{equation}%\lablel{eq12.1}
\begin{split}
1-STEP(Eviction_Moratorium_ON*\\
Magnitude_Reduction_in_Filings_During_Moratorium, TS_EM-.5) \\
+SMTH1(STEP(Eviction_Moratorium_ON*\\
Magnitude_Reduction_in_Filings_During_Moratorium, TS_EM+Duration_EM+3), 36\\
Where: 	ES_EM = Effect size of eviction moratorium\\
TS_EM = Start time of eviction moratorium\\
ET_EM = End time of moratorium
\end{split}
\end{equation}

Second, the Consolidated Appropriations Act of 2021 and the American Rescue Plan together allocated nearly \$50 billion in emergency rental assistance (ERA) for low-income renters impacted by the pandemic. A stock-and-flow structure reflecting the initial available ERA funds and estimated rate of disbursement by the U.S. Treasury \cite{national_low_income_housing_coalition_treasury_2022} was linked to rent payments to reflect assistance filling in gaps for delinquent tenants (Figure 12). 

\begin{figure} % picture
    \centering
    \includegraphics{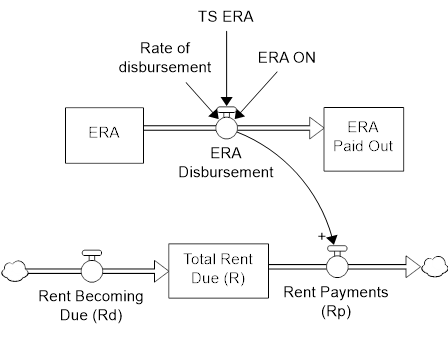}
     \caption{Emergency rental assistance structure} 
\end{figure}

\subsection{Model Calibration and Validation}
The model was calibrated using data from a range of sources. Estimates of literal homelessness came from the 2018-2020 Annual Homeless Assessment Reports (AHAR), which include point-in-time counts every January of sheltered and unsheltered homelessness \cite{us_census_bureau_american_2018,joint_center_for_housing_studies_of_harvard_university_americas_2020,joint_center_for_housing_studies_of_harvard_university_state_2020}, and the National Low-Income Housing Coalition \cite{national_low_income_housing_coalition_renters_2012,national_low_income_housing_coalition_gap_2021}. Estimates of eviction filings and evictions came from Eviction Lab \cite{desmond_eviction_2018, eviction_lab_eviction_2022}. Finally, the impact of COVID-19 on low-income rental housing market was calibrated based on unemployment rates and self-reported economic hardship \cite{federal_reserve_bank_of_st_louis_unemployment_2021, us_census_bureau_household_2021}. When available, historical data were plotted to create reference modes; model parameters were adjusted with empirical and theoretical justification so that simulations were able to replicate reference modes. 

Uncertainty underlay the model structure and reference modes given imperfect existing data and the sensitive nature of housing decisions among very low-income households. Thus a range of prior studies contributed empirical and theoretical knowledge to the model-building process \cite{eskinasi_houdini_2011,fowler_solving_2019,marcal_understanding_2021,marzouk_modeling_2016}. Furthermore, several measures built confidence in model structure and simulation results. Model simulated trends were compared to historical data using Theil’s U, an indicator of bias and variation in prediction (Appendix III). Parameter sensitivity analyses simulated the model across +/- 15\% of all initial values to test robustness of the model structure to initial conditions. Inputting extreme parameters assessed model behavior under unrealistic \cite{saleh_comprehensive_2010, senge_tests_1980}. Shapes of graphical functions were first built according to prior theoretical and empirical research; algebraic equations underlying functions were used to generate parameters for sensitivity analyses in R Version 4.0.3 \cite{oliva_model_2003}. Key housing policy experts consulted on model structure and outputs in an iterative fashion as an additional validity check. All model building and simulations were conducted in Stella Architect Version 2.1.

Once the model was calibrated, simulations reflected alternate scenarios for policy evaluation and planning. Run 1 reflected pre-pandemic dynamics in the low-income rental market and assess the accuracy of model outputs relative to real-world data. Run 2 included the economic impact of COVID-19 on the low-income rental housing market with no policy interventions. Run 3 tested the impact of COVID-19 with the federal eviction moratorium in place for 18 months \cite{mccarty_federal_2021, national_low_income_housing_coalition_national_2021}. Run 4 added the federal Emergency Rental Assistance (ERA) program, which allocated \$46.5 billion to keep tenants housed.

The total time horizon for simulations was 50 months (January 2018 through February 2022). This included a 24-month “burn-in” period such that Month 1 represented January 2020 and the analytical time horizon was 26 months – through February 2022. The model was simulated using the Euler integration method with a delta time of 0.25 months.

\subsection{Simulation Results}
Simulations estimate the economic fallout from COVID-19 severely impedes low-income households’ ability to afford housing without intervention in Figure 13 and Figure 14. Tenants accumulate \$20.4 billion in unpaid rent over 36 months, which leads to 1.5 million excess evictions across the follow-up – an increase of over 25\% compared to pre-pandemic rates. Unpaid rent also drives crowding and homelessness, which increase 45\% and 120\% respectively. 

The federal eviction moratorium enacted in March 2020 dramatically reduces tenant displacements through September 2021 compared with no federal intervention. The moratorium removes incentives for tenants to double up to avoid housing cost burden and eviction, however; with the moratorium in place, arrears remain at nearly \$20 billion by the end of the simulation period. Eviction filings lagged in the wake of the moratorium’s expiration \cite{eviction_lab_eviction_2022}, keeping eviction rates below pre-pandemic rates through the follow-up period despite signs of an initial rebound. The moratorium reduces total evictions by 51\% across the 26-month follow-up compared to no intervention. Pausing evictions allows some households time to pay off debts, double up, or move before facing formal displacement.

Federal emergency rental assistance further aids households in retaining housing in spite of the COVID-19 economic fallout. Initial roll-out was delayed, with states slow to disburse funds due to complicated eligibility criteria and application processes for both tenants and landlords. As of February 2022, only 42\% of funds allocated by the federal government had been paid out to tenants. Simulation results show that with this rate of disbursement, arrears, crowding, and homelessness remain high despite \$27 billion in assistance remaining unused. An additional simulation (Run 4a) builds on this finding to tests an alternate scenario wherein ERA funds are disbursed more rapidly from the start; under these conditions, arrears, crowding, and homelessness plummet quickly. Although funds run out sooner, each dollar goes further by stabilizing cost-burdened and homeless households more efficiently. Increasing speed of disbursement burns through funds more quickly but heads off additional accumulation of arrears that feed enduring housing insecurity, crowding, and homelessness. 

\begin{figure} % picture
    \centering
    \includegraphics{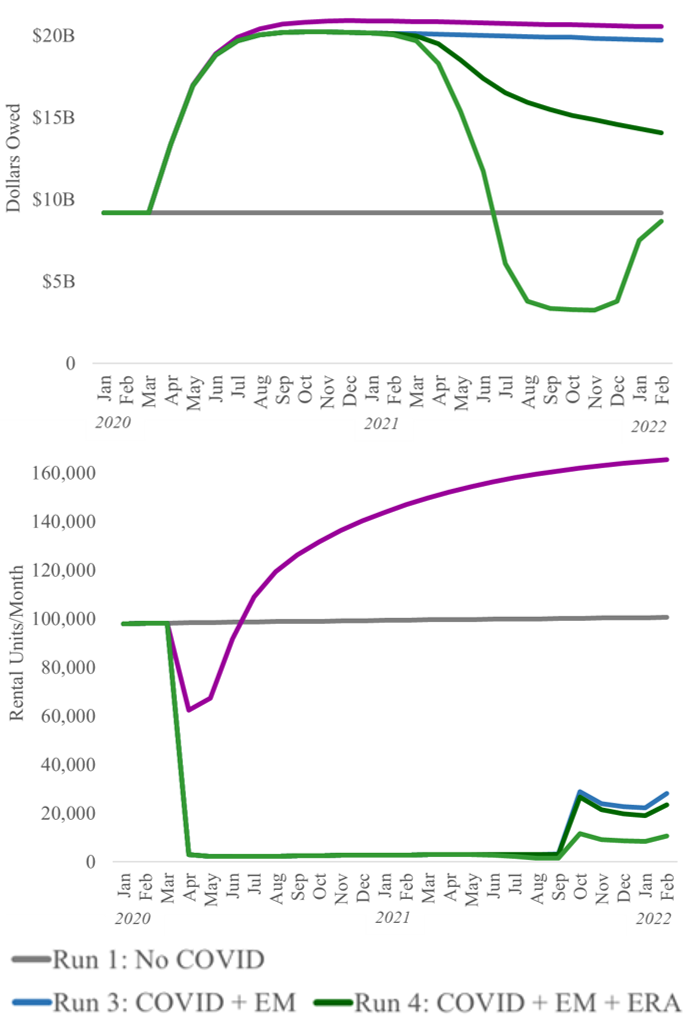}
     \caption{Results of simulations testing the impact of COVID-19, the federal eviction moratorium (EM), and federal emergency rental assistance (ERA) on accumulated arrears (top) and monthly evictions (bottom)}
\end{figure}

\begin{figure} % picture
    \centering
    \includegraphics{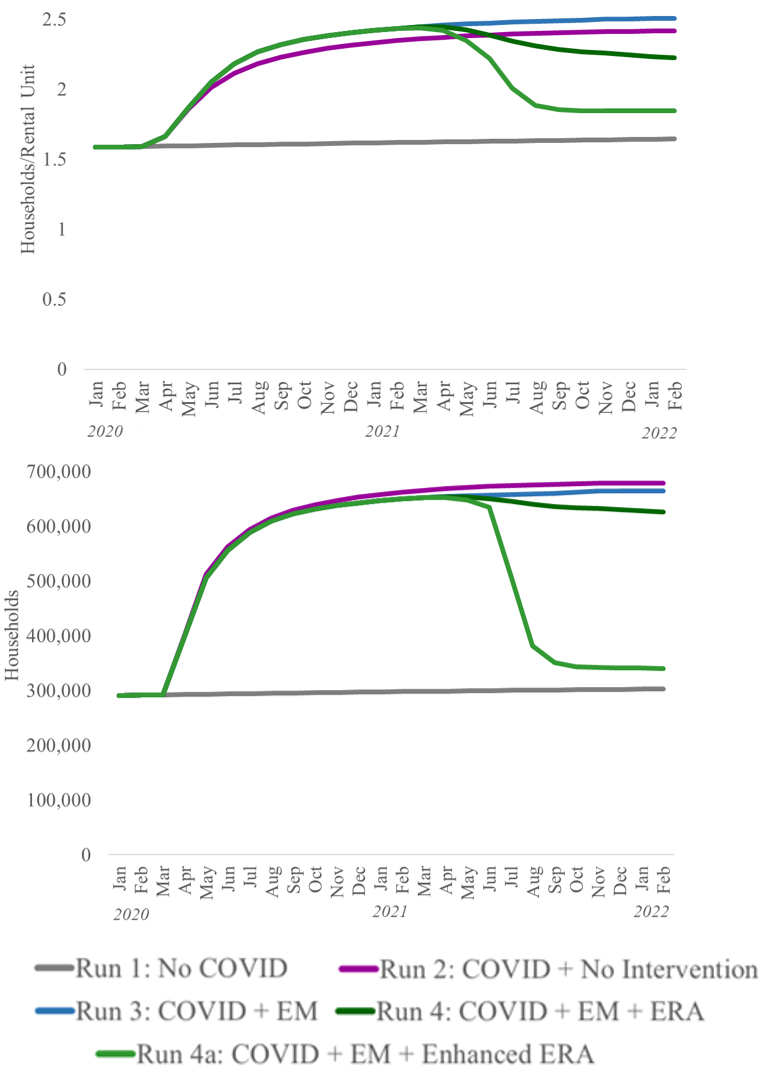}
    \caption{Results of simulations testing the impact of COVID-19, the federal eviction moratorium (EM), and federal emergency rental assistance (ERA) on average crowding (top) and literal homelessness (bottom)} 
\end{figure}

\section{Discussion}
The low-income rental housing market is marked by instability, scarcity, and difficult tradeoffs. In the present study, a system dynamics model highlights the complex dynamics driving tenant and landlord behaviors in the context of limited affordable housing and precarious incomes. Prior to COVID-19, vicious cycles of household economic stress and interpersonal conflict hinder tenants’ abilities to secure and maintain stable housing; compensatory processes lead landlords to calibrate to market conditions in order to keep units occupied and avoid foreclosures. These feedback mechanisms underlie persistent rates of housing insecurity and homelessness over time. The economic fallout from the pandemic exacerbates low-income rental market vulnerabilities. Pervasive economic hardship in the absence of widely available supports fuels the low-income population with high barriers to housing. Post-COVID evictions displace tenants, exposing households to overcrowded or unsafe conditions as well as literal homelessness. Low-income landlords continuously consider how to pay their mortgages and make a profit. Eviction decisions depend on tenant delinquency as well as landlords’ own financial hardship. The COVID-19 recession and the policy responses introduce volatility into a low-income housing system driven by complex decision-making processes.

Simulations suggest trends in evictions and literal homelessness only partially reflect instability and hardship that characterize the low-income rental market. Widespread financial strain as households accumulate arrears triggers decisions aimed at avoiding eviction and homelessness – two of the most visible, easily tracked outcomes. Doubling up allows more families to avoid homelessness, but contributes to conflict that may erode living conditions over time. The federal eviction moratorium implemented in response to COVID-19 paused formal evictions while household economic strain persisted; projections suggest this translated to increased crowding and homelessness with delays. While tenants aim to avoid evictions, landlords rely on evictions to minimize lost income. Future efforts to stabilize housing and reduce homelessness for low-income households must consider decision-making in the context of these conflicting goals. The eviction moratorium may have had additional benefits beyond the scope of the present study. While the lowest-income renters typically face the highest risk for eviction and thus were most likely to benefit from the eviction moratorium, the moratorium applied equally to all renters and thus likely kept some higher-earning tenants in housing and out of the model population. In general, the eviction moratorium would have allowed renters regardless of income level to remain in their current housing. Findings show lasting benefits of the moratorium for evictions, but arrears, crowding, and homelessness do not improve without significant government financial investment for the most vulnerable renters.

Parameter sensitivity analyses tested the robustness of the model structure to initial parameter values +/-15\%. The ratio of total rental units to low-income households emerges as an area of sensitivity; the low-income rental market relies on a certain level of doubling up, such that combining incomes allows more families to stay housed in the mechanism described by B1. When the number of rental units increases relative to households such that doubling up decreases, evictions and homelessness spike; this dynamic points to the challenges of increasing affordable housing that may not actually be affordable for very low-income households juggling multiple monthly expenses that strain resources.

A number of policy implications arise from the present study. Housing policies must offer opportunities to avoid eviction filings that benefit both landlords as well as tenants. The eviction moratorium demonstrates that pausing or delaying evictions of low-income households can lead to permanent declines as tenants have more opportunities to catch up on payments or find alternate living arrangements without experiencing formal eviction or entering homelessness. Measures that slow the eviction process in the low-income rental market incentivize landlords to work with tenants on payment plans rather than turning to the courts. Although a need exists for broader economic and housing policies that make housing more affordable, slowing evictions stabilize severely housing-burdened households without triggering unintended consequences, including raised rents and gentrification \cite{ellen_how_2010}. 

A number of feasible options exist for slowing evictions. One potential longer-term solution would be to reform the eviction policies to mirror the foreclosure process. Currently, landlords are protected by federal law that prohibits lenders from initiating foreclosures in most cases until the borrower is over 120 days delinquent (12 C.F.R. § 1024.41). In contrast, evictions occur quickly; a recent study found that the average debt owed by tenants in eviction hearings was \$1,200 \cite{urban_cleveland_2019} or just over one months’ rent at the U.S. median level \cite{us_census_bureau_american_2018}. Nationally, 60\% of eviction judgements are decided in the plaintiff’s favor \cite{desmond_eviction_2018}. A slower eviction process makes filing evictions a less attractive option for landlords, who may increase their tolerance for overdue rent and be more willing to accommodate payment plans. Second, removing evictions from tenant records after a period of time may mitigate the effect of evictions on future housing stability, allowing tenants to more easily find housing and landlords to more easily fill vacant units. Imposing time limits on the impact of eviction histories can prevent erosion of renter applicant quality and more quickly enable households to exit homelessness. Finally, rapid, widespread dispersion of rental assistance allocated as part of the federal COVID response can address the outstanding arrears that contribute to ongoing financial strain despite the eviction moratorium. Despite billions of dollars allocated to states for rental assistance, technological limitations and complex eligibility criteria have hindered efficient distribution of funds to households in need. More rapid payments can keep low-income families housed and mitigate housing insecurity as the eviction moratorium lifts.

\subsection{Limitations}
Findings from modeling must be considered in light of study limitations. First, the model is calibrated at a national level using the best available data; results thus provide an aggregate view of the U.S. low-income rental market, but do not capture geographic variation in rental markets due to differing costs of living, affordable housing availability, and labor markets. Future modeling should consider local variation to estimate trends in specific metropolitan areas or geographic regions. Second, results reflect assumptions about sustained economic and public health recovery from the COVID-19 pandemic; future surges driven by variants or stalled vaccination rates may drive future mitigation efforts that prolong or alter the economic recovery. 

The model also does not track cumulative harm to households from experiencing eviction, homelessness, and crowding. In most states, eviction histories impact credit scores for years, impeding the ability to secure new housing. Exposure to literal homelessness is associated with enduring stigma and adverse consequences for physical and mental health. Prolonged experiences of inadequate housing, eviction, and homelessness may also drive households to accrue additional late fees or seek emergency loans through payday lenders, further impacting financial hardship over time. Future modeling should consider the potential long-term, cumulative effects of housing insecurity on tenant behaviors and outcomes. 

Finally, the pandemic-triggered economic impact and recovery are modeled as exogenous variables unaffected by trends in the low-income rental market and policy responses to the affordable housing crisis. Examination of the interactions between the affordable housing market and broader economic recovery is an important area of future research.	

\subsection{Implications for Future Research}
Despite limitations, the present study highlights important insights into the dynamics underlying the low-income rental market that can inform post-COVID relief efforts. The model points to a number of important opportunities for future inquiry. First, little is known about complex landlord decision-making in response to market conditions. In slack markets, landlords may adjust expectations for tenants and tolerate late rent for longer before initiating eviction filings with the assumption that late rent is better than no rent \cite{balzarini_working_2021}; landlords may also respond differently to eviction histories when evictions are widespread and it becomes difficult to find tenants with clean records. The dynamics underlying landlord cost-benefit analyses should be explored using participatory approaches and incorporated into future model iterations. Second, low-income households make complex tradeoffs in household budgeting to juggle unaffordable basic needs \cite{beatty_is_2014}. A thorough accounting of how household expenses are managed when incomes are too low to pay all bills each month may inform efforts to provide assistance during economic downturns. Finally, delays in disbursing ERA funds have impeded efforts to stabilize the low-income rental market; modeling efforts should examine barriers to rapid implementation of assistance during times of crisis. Future research should evaluate efforts to address widespread financial hardship that reduce emphasis on individual eligibility criteria and focus on systemic interventions \cite{ellen_how_2010}.

\subsection{Conclusions}
The expiration of the eviction moratorium and delayed rollout of emergency rental assistance threatens a continuing crisis of housing insecurity and homelessness in the wake of the COVID-19 pandemic. Understanding the tradeoffs and conflicting interests of tenants and landlords in a context of scarcity can help target interventions that keep families housed. Rapid assistance that addresses accumulated arrears and reduces incentives for landlords to evict is necessary for housing security and public health in vulnerable communities.

%\addbibresource{mybibliography.bib}
%\printbibliography
%\nocite{*}
\bibliographystyle{unsrt}
\bibliography{bibliography.bib}

%\newpage

%\appendix
%\appendixpage

%\begin{appendices}
%\end{appendices}

\end{document}